\documentclass[%
 reprint,
superscriptaddress,
showpacs,preprintnumbers,
 amsmath,amssymb,
 aps,
]{revtex4-1}

\usepackage{graphicx}
\usepackage{dcolumn}
\usepackage{bm}

\usepackage[caption=false]{subfig}
\usepackage{epstopdf} 

\newcommand\redout{\bgroup\markoverwith{\textcolor{red}{\rule[.5ex]{2pt}{0.4pt}}}\ULon}

\usepackage[colorlinks=true,citecolor=blue]{hyperref}
\hypersetup{colorlinks=true,citecolor=blue,linkcolor=red,urlcolor=blue}

\begin{document}
\title{Many-body exchange-correlation effects in MoS$_2$ monolayer:
 The key role of nonlocal dielectric screening}
\author{A. Faridi}
\email{azadeh.faridi@ipm.ir}
\affiliation{School of Physics, Institute for Research in Fundamental Sciences (IPM), Tehran 19395-5531, Iran}

\author{Reza Asgari}
\affiliation{School of Physics, Institute for Research in Fundamental Sciences (IPM), Tehran 19395-5531, Iran}
\affiliation{School of Nano Science, Institute for Research in Fundamental Sciences (IPM), Tehran 19395-5531, Iran}
\affiliation{ARC Centre of Excellence in Future Low-Energy Electronics Technologies, UNSW Node, Sydney 2052, Australia}

\date{\today}

\begin{abstract}
We calculate the quasiparticle properties of $\rm{MoS_2}$ monolayer at $T=0$ considering the dynamical electron-electron interaction effect within random-phase-approximation (RPA). The calculations are carried out for an electron-doped slab of $\rm{MoS_2}$ monolayer using a minimal massive Dirac Hamiltonian and the quasi-two-dimensional nature of the Coulomb interaction in this system is taken into account considering a modified interaction of Keldysh type. Having calculated the real and imaginary parts of the retarded self-energy, we find the spectral function and discuss the impact of extrinsic variables such as the dielectric medium and the charge carrier density on the appearance and position of the quasiparticle peaks. We also report the results of the renormalization constant and the effective Fermi velocity calculations in a broad range of the coupling constant and carrier density. We show that the effective Fermi velocity obtained solving the self-consistent Dyson equation has an absolutely different behavior from the one found from the on-shell approximation. Our results show that the nonlocal dielectric screening of the monolayer tends to stabilize the Fermi liquid picture in $\rm{MoS_2}$ monolayer and that the interaction strength parameter of this system is a multivariable function of the coupling constant, carrier density, and also the screening length.
\end{abstract}


\maketitle

\section{Introduction}
Following the growing interest in new two-dimensional materials inspired by the discovery of graphene, monolayer of molybdenum disulfide ($\rm{MoS_2}$), a prototypical member of transition metal dichalcogenides (TMDs), has attracted a great deal of attention for about fifteen years due to its distinguished electronic and optoelectronic properties such as high mobility \cite{radisavljevic2011single}, valley Hall effect \cite{mak2014valley}, strong photoluminescence \cite{splendiani2010emerging}, and emergence of tightly bound neutral and charged excitons \cite{berkelbach2013theory}. Composed of a hexagonal plane of molybdenum atoms sandwiched between two hexagonal layers of sulfur atoms, $\rm{MoS_2}$ monolayer is a direct band gap semiconductor \cite{mak2010atomically} in contrast to its indirect gap bulk counterpart \cite{boker2001band} which has been known for almost five decades. 
The sizable band gap of $\rm{MoS_2}$ located at $\rm{K,K'}$ points in the Brillouin zone ranges from visible to IR \cite{liu2019direct}, and it is the privileged feature of $\rm{MoS_2}$ in comparison with graphene which makes it appropriate for electronic and optoelectronic-based technologies.

A great deal of research has been conducted in the past couple of years in order to reveal the interesting and peculiar characterizations of $\rm{MoS_2}$ among which the optoelectronic and optical studies have received much consideration \cite{ramasubramaniam2012large,wu2015exciton,yin2014edge,ben2016optoelectronic}. Many groups have also worked on the electronic and quasiparticle properties of $\rm{MoS_2}$ both theoretically and experimentally \cite{huser2013dielectric,qiu2016screening,huser2013dielectric,van2019probing,thygesen2017calculating,PhysRevB.91.235301}. One of the appealing features of $\rm{MoS_2}$ and TMDs in general which motivates a deeper understanding of the electronic properties, is their external controllability of the quantum many-body properties. In particular, the environmental sensitivity of the Coulomb interaction in these materials is remarkable. It has been shown that the dielectric feature of the surrounding medium has influential effects on the many-body screening and thereby on the electronic properties and enables us to engineer some quasiparticle characterizations \cite{raja2017coulomb}.
On the other hand, doping as a typical task in two-dimensional (2D) semiconductors has crucial consequences in the determination of the quasiparticle properties through further screening effects caused by doped carriers. This effect has also been discussed in the case of $\rm{MoS_2}$  \cite{yao2017optically}.

Accordingly, it became evident that at this point a systematic investigation of the quantum many-body effects in this system can be of interest from both fundamental and application aspects. The main purpose of this article is to present a theoretical analysis in order to determine a comprehensive picture of the quasiparticle features of  $\rm{MoS_2}$ and also of the extrinsic variables affecting them. In doing so, we carry out a full random-phase-approximation (RPA) self-energy calculation in the framework of Landau Fermi liquid theory for an electron-doped (no photoexcited carriers) slab of $\rm{MoS_2}$ monolayer.

The well-established RPA self-energy formulation used in this paper was first discussed by Quinn and Ferrell in their celebrated paper in order to find the correlation energy of a degenerate electron gas \cite{quinn1958electron}. Since then, the method has been widely used by many authors to describe the quasiparticle-quasiparticle interactions in several electronic systems ranging from metals to semiconductors in all dimensions and in a broad range of interaction strengths. Using this method, Lundqvist calculated the single-particle spectrum of a three-dimensional electron gas in the range of metallic densities and he found an extra low-energy peak (plasmaron) corresponding to the plasmon-hole coupling aside from the typical quasiparticle peak in the spectral function \cite{lundqvist1967single,lundqvist1968single,hedin1970effects}. Similar studies were also performed widely for 2D and quasi-2D electron gas after its realization in II-VI and III-V semiconductor heterostructures \cite{asgari2005quasiparticle,von1992plasmaron,giuliani1982lifetime,jalabert1989quasiparticle}. The method was also used for many other individual systems such as quasi one-dimensional electronic systems known as quantum wires \cite{hu1993many} and ultracold dipolar Fermi liquids \cite{seydi2018composite} to name a few. In recent years, the RPA self-energy calculations have been successful in theoretical description of the quasiparticle properties and spectral function of graphene and Dirac materials \cite{polini2008plasmons,bostwick2010observation,hwang2008quasiparticle,qaiumzadeh2009effect}. The same calculations are done here in order to precisely describe the quantum many-body effects in $\rm{MoS_2}$ monolayer. In particular, in this work, we consider the important quasi-2D nature of the $\rm{MoS_2}$ monolayer using a modified Coulomb interaction with a nonlocal momentum-dependent dielectric function. In short, the contributions of the current study are: (1) Evaluation of some many-body properties of $\rm{MoS_2}$ monolayer using a modified Coulomb interaction such as quasiparticle energy, spectral function, renormalization constant and renormalized Fermi velocity. (2) Exploring the impact of external variables namely doped carrier density, surrounding medium, and screening length on the quasiparticle features of the system.

The paper is organized as follows.
In Sec. II we focus on preliminaries and theoretical structure we have used in our calculations and introduce the model Hamiltonian for $\rm{MoS_2}$, the quasi-2D Coulomb interaction, and the perturbative self-energy formalism within RPA. In Sec. III we define some quantum many-body properties of Fermi liquids and present our main numerical results of the real and imaginary parts of the self-energy, the spectral function, the renormalization constant, and the effective Fermi velocity in a broad range of density and coupling constant. Section IV contains our summary and conclusions.

\section{Theory and Formalism}
In this section, we  briefly present the theoretical assumptions and framework of our paper including the low-energy model Hamiltonian and the Keldysh Coulomb potential as well as the quasiparticle self-energy formalism. 

\subsection{Effective low-energy Hamiltonian}
We consider an electron-doped monolayer of $\rm{MoS_2}$ on a substrate with long-range electron-electron interaction. Neglecting the intervalley scattering the total Hamiltonian at K point is given by
\begin{equation}
 \begin{split}\label{eq:fullH1}
{\hat {\cal H}} &= \sum_{ k,\sigma}{\hat \psi}^\dagger_{\bm k,\sigma}(a_0t \bm k\cdot\hat{\bm{\sigma}}+\Delta\hat{\sigma_z}-\sigma\lambda\frac{\hat{\sigma}_z-1}{2}){\hat \psi}_{\bm k,\sigma}\\
&+\frac{1}{2 S}\sum_{{\bm q} \neq {\bm 0}} V(q) {\hat \rho}_{\bm q} {\hat \rho}_{-{\bm q}},
\end{split}
\end{equation}
where ${\hat \psi}^\dagger_{\bm k,\sigma}=(\hat a^\dagger_{\bm k,+\sigma},\hat b^\dagger_{\bm k,+\sigma})$ is the pseudospin operator, ${\hat \rho}_{\bm q} = \sum_{{\bm k},\sigma} {\hat \psi}^\dagger_{{\bm k} - {\bm q},  \sigma}{\hat \psi}_{{\bm k}, \sigma}$ is the density operator, and $V(q)$ is the quasi-2D electron-electron interaction.The first term is the noninteracting minimal two band Hamiltonian of the massive Dirac fermions proposed for the monolayer of $\rm{MoS_2}$ which is written in lowest order $k\cdot p$ theory \cite{PhysRevB.88.085440,xiao2012coupled}.
Here $\hat{\bm{\sigma}}$ denotes the Pauli matrices acting in pseudospin space, $\sigma$ is the real spin, $t=1.10\, \rm {eV}$ is the hopping matrix element, $a_0=3.193$ \AA~is the lattice constant, $2\Delta=2.7 \, \rm{eV}$ is the electronic energy gap between the valence and conduction bands \cite{yao2017optically,gao2017renormalization}, and
 $2\lambda=0.15 \,\rm{eV}$ is the spin splitting of the valence band.
Having an electron-doped system, we thus ignore this term for the sake of convenience, since it is much smaller than the electronic band gap and has minor effect on the quasiparticle properties of the system ($\lambda\approx0.06\Delta$). The eigenvalues of this noninteracting term of the Hamiltonian are given by ${E_ k^s=s\sqrt{(\hbar v_F k)^2+\Delta^2}}$  where $v_{\rm F}=a_0t/\hbar\approx5.33\times10^{5}~\rm{m/s}$ is the Fermi velocity and $s=+/-$ denotes the conduction/valence band. The second part of Eq.~\eqref{eq:fullH1} is the long-range Coulomb interaction which will be treated perturbatively.

\subsection{Quasi-2D Coulomb interaction}
In ordinary 3D materials the effect of lattice screening is simply a rescaling of the interaction strength by a static dielectric constant. In 2D materials with finite width, however, the nonlocal dielectric screening leads to the modified Coulomb interaction of the form \cite{cudazzo2011dielectric,torbatian2018plasmonic} 
\begin{equation}\label{vq}
V(q,a)=\frac{2\pi e^2}{\epsilon(q+aq^2)},
\end{equation}
where $a$ is related to the polarizability of the 2D layer through $a=2\pi\alpha_{\rm{2D}}$~\cite{cudazzo2011dielectric} and $\epsilon=(\epsilon_1+\epsilon_2)/2$ is the average dielectric constant of the environment.
The Fourier transform of this interaction is no longer $e^2/r$, however, it is

\begin{equation}\label{vr}
V(r,a)=\frac{e^2\pi [-Y_0(r/a) + H_0(r/a)]}{2a\epsilon},
\end{equation}
where the Bessel function of the second kind is defined as
\begin{equation}
Y_n(x)=\frac{J_n(x)\cos(n x)-J_{-n}(x)}{\sin(nx)},
\end{equation}
where $J_n(x)$ is the Bessel function of the first kind and the Struve function $H_n(x)$ solves the inhomogeneous Bessel equation. It can be shown that for $r\to \infty$ the interaction reduces to the formal Coulomb interaction and it shows a weaker logarithmic divergence as $r\to 0$. This potential was first proposed in the Keldysh model~\cite{keldysh1979coulomb} for a geometry in which a slab of thickness $d$ and isotropic in-plane dielectric constant $\epsilon_{\parallel}$ is assumed to be sandwiched between materials with dielectric constants $\epsilon_1$ and $\epsilon_2$.  In this model the screening length is approximated as $a=d\epsilon_{\parallel}/(\epsilon_1+\epsilon_2)$.
 For an in-plane dielectric constant of $\epsilon\approx12$ and a slab thickness of $d\approx6$, Zhang $\it et~ al$~\cite{zhang2014absorption} found that the screening length is $a\approx36$ \AA~for a freestanding MoS$_2$ monolayer ($a=36/\epsilon$~\AA~in general). This is in good agreement with $a=35$~\AA~ found by Qiu $\it et~ al$~\cite{qiu2016screening} fitting the Keldysh model to their {\it ab initio} effective dielectric function at small $q$. 

\subsection{Many-body self-energy within RPA}
In order to find quasiparticle properties of an interacting system we should have information of the Green's function or equivalently from the self-energy of the system. Here we have used the  G$^0$W self-energy which is based on two main approximations: First, the self-energy is written in leading order in the dynamical interaction and the vertex corrections are neglected and second, the interacting Green's function of the system $G$ is replaced by the noninteracting one $G^0$. In this regard, at zero temperature, $T=0$, the retarded self-energy $\Sigma_s$ of the quasiparticles in the conduction band ($s\to+$) or the valence band ($s\to-$) is given by \cite{mahan2013many,giuliani2005quantum}
\begin{equation}\label{self1}
 \begin{split}
&\Sigma_s(\bm k,\omega)=  \\
&-\sum_{s'}\int\frac{d^2\bm q}{(2\pi)^2}F^{ss'}_{\bm k,\bm k+\bm q}\int_{-\infty}^{\infty}\frac{d\Omega}{2\pi i}\frac{V_q}{\epsilon(\bm q,\Omega)}G^0_{s'}(\bm k+\bm q,\omega+\Omega),
 \end{split}
\end{equation}
where $V_q$ is the short form of $V(q,a)$ given by Eq.~\eqref{vq} and ${\epsilon(\bm q,\Omega)=1-V_q\chi^0(q,\Omega)}$ is the dynamical dielectric function within RPA and $\chi^0(\bm q,\Omega)$ is the noninteracting polarization function of the system \cite{pyatkovskiy2008dynamical}. $F^{ss'}_{\bm k,\bm k+\bm q}$ is the wave function overlap factor of the states $s$ and $s'$ and is given by \cite{qaiumzadeh2009effect} 
\begin{equation}
F^{ss'}_{\bm k,\bm k+\bm q}=\frac{1}{2}(1+ss'\frac{\hbar^2v_F^2(\bm k\cdot \bm{k}+\bm{q})+\Delta^2}{E_{\bm k}E_{\bm{k+q}}}).
\end{equation}
The noninteracting Green's function of the system is defined as
\begin{equation}
G^0_{s}(\bm k,\omega)=\frac{1-n_F(\xi_{\bm k}^s)}{\omega-\xi_{\bm k}^s+i\eta}+\frac{n_F(\xi_{\bm k}^s)}{\omega-\xi_{\bm k}^s-i\eta},
\end{equation}
where $\eta$ is an infinitesimal positive constant and ${\xi_{\bm k}^s=E_{\bm k}^s-E_F}$ is the noninteracting energy measured from Fermi level. Here we assume an electron-doped system with the Fermi energy $E_F=\sqrt{(\hbar v_F k_F)^2+\Delta^2}$. $n_F(\xi_{\bm k}^s)$ is the Fermi distribution function and at $T=0$ we have $n_F(\xi_{\bm k}^s)=\Theta(-\xi_{\bm k}^s)$ with $\Theta(x)$ being the Heaviside function. In this work, we consider an experimentally accessible density range (up to  $n=\rm{5\times10^{13}~cm^{-2}}$) for which the Fermi energy does not exceed $1.42~ \rm{eV}$ showing the electron gas is confined at the bottom of the conduction band. Increasing the carrier density to much larger values, the system can get closer to a gapless Dirac system, but the density range in this case is experimentally inaccessible. 

The retarded self-energy in Eq.~\eqref{self1} can be decomposed into the static exchange part $\Sigma_s^{\rm ex}$ and the dynamical correlation part $\Sigma_s^{\rm cor}$ 
\begin{equation}
\Sigma_s(\bm k,\omega) =\Sigma_s^{\rm{ex}}(\bm k) +\Sigma_s^{\rm cor}(\bm k,\omega).
\end{equation}
The exchange self-energy is simply given by
\begin{equation}\label{ex}
\Sigma_s^{\rm{ex}}(\bm k,\omega)= -\sum_{s'}\int\frac{d^2\bm q}{(2\pi)^2}V_q F^{ss'}_{\bm{k},\bm k+\bm q} \Theta(-\xi^{s'}_{\bm k +\bm q}).
\end{equation}
The $\Omega$ - integration on the real axis  in the correlation part of Eq.~\eqref{self1} encounters some difficulties owing to the poles of $1/\epsilon(\bm q,\Omega)$. This problem is avoided by closing the integration contour in the first and third quadrants using two circular contours. Then we are left with the sum of the Green's function residues in the first and third quadrants plus an integration on the imaginary axis \cite{rice1965effects}. Note that the circular contours do not contribute to the integration because of the decaying behavior of the integrand at $\Omega\to\infty$. Following these steps, the correlation part of the self-energy can be written as the sum of a line and a residue term 
\begin{equation}
\Sigma_s^{\rm cor}(\bm k,\omega) =\Sigma_s^{\rm{line}}(\bm k,\omega) +\Sigma_s^{\rm res}(\bm k,\omega) ,
\end{equation}
where
\begin{equation}\label{line1}
\begin{split}
&\Sigma_s^{\rm line}(\bm k,\omega)=\\
&-\negthickspace\sum_{s'}\negthickspace\int\negthickspace\frac{d^2\bm q}{(2\pi)^2}V_qF^{ss'}_{\bm k,\bm k+\bm q}\negthickspace\int_{-\infty}^{\infty}\negthickspace\frac{d\Omega}{2\pi}\Bigl[\frac{1}{\epsilon(\bm q,i\Omega)}\negthickspace-1\negmedspace\Bigr]\frac{1}{\omega\negmedspace+\negmedspace i\Omega\negmedspace-\negmedspace\xi^{s'}_{\bm k+\bm q}},
\end{split}
\end{equation}
and 
\begin{equation}\label{res}
\begin{split}
&\Sigma_s^{\rm res}(\bm k,\omega)=\\
&\sum_{s'}\int\frac{d^2\bm q}{(2\pi)^2}V_q\Bigl[\frac{1}{\epsilon(\bm q,\omega-\xi_{s'}(\bm k+\bm q))}-1\Bigr]F^{ss'}_{\bm k,\bm k+\bm q}\\
&\qquad\times\bigr[\Theta(\omega-\xi^{s'}_{\bm k+\bm q})-\Theta(-\xi^{s'}_{\bm k+\bm q})\bigl].
\end{split}
\end{equation}\vspace{3mm}

We can see that the line contribution to the correlation self-energy is completely real since 
$\epsilon(\bm q,i \Omega)$ is a real quantity. Therefore, the only contribution to the imaginary part of the self-energy comes from the residue term.
\begin{figure}[h]
\centering
\subfloat{%
  \includegraphics[width=1.\linewidth]{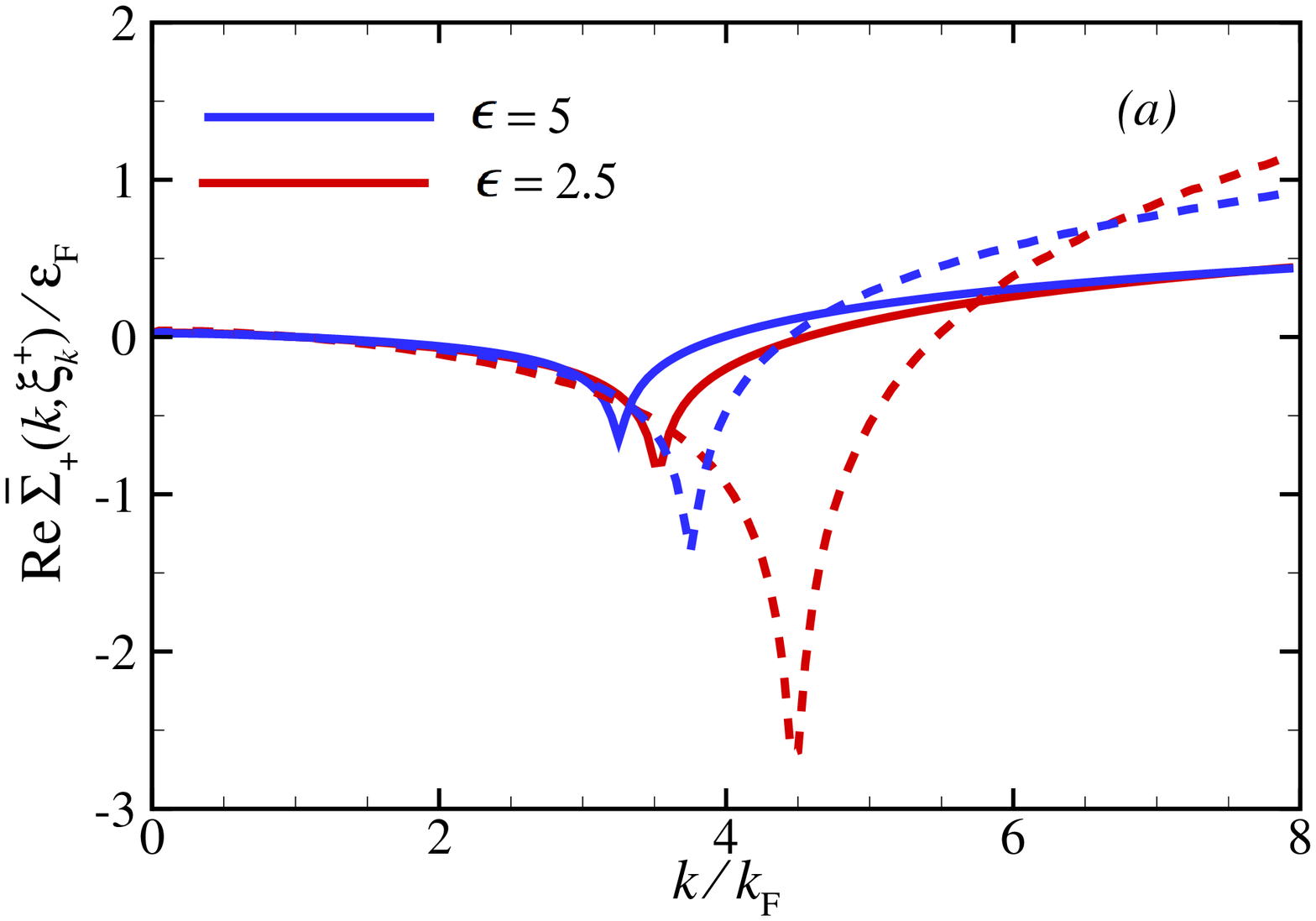}%
  }\par 
\negthickspace\negthickspace\negthickspace\negthickspace
\subfloat{%
  \includegraphics[width=1.\linewidth]{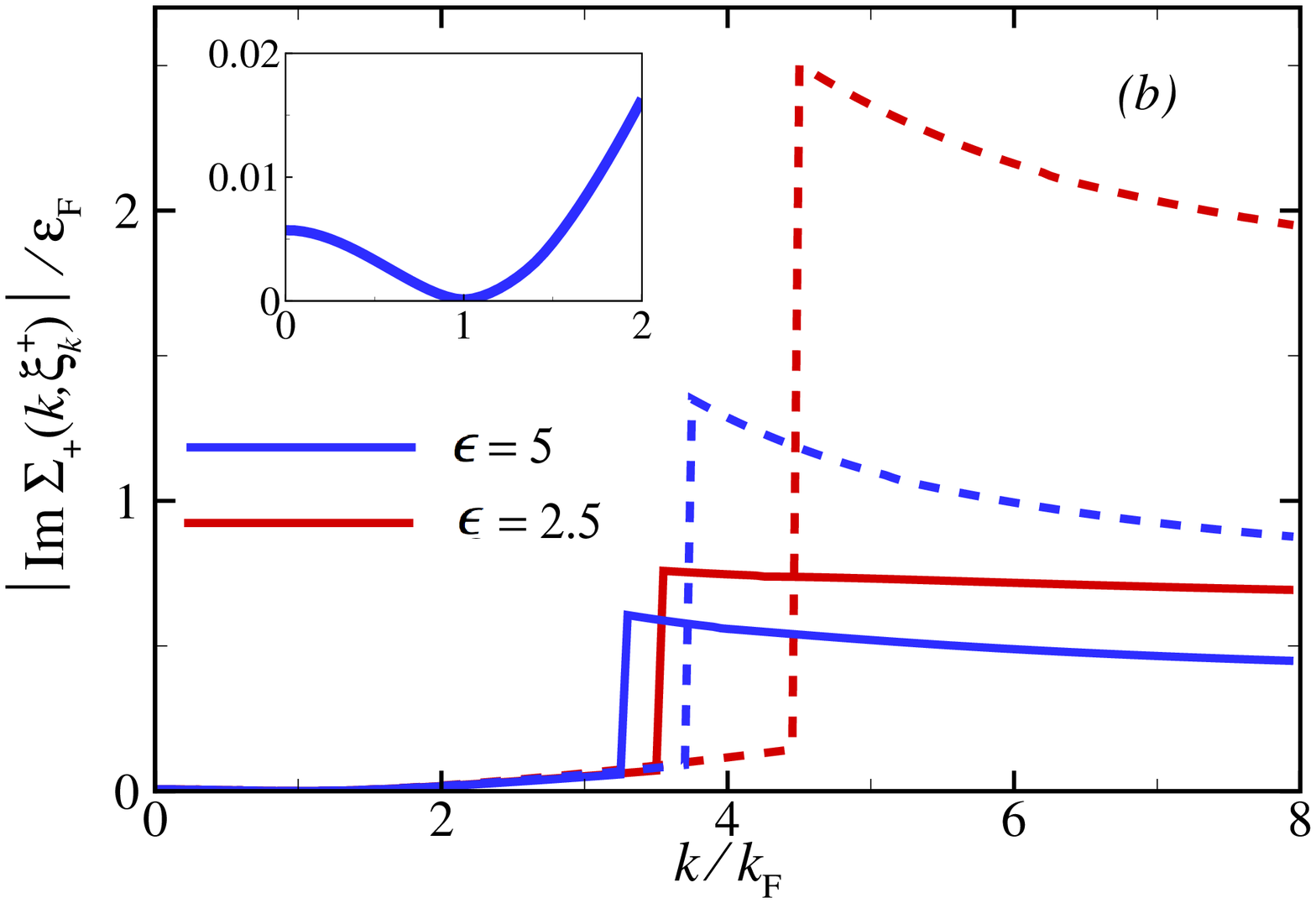}%
  }
\caption{\label{fig1} (Color online) (a) The real part of the self-energy with respect to the interacting chemical potential $\mathrm{Re}\,\bar{\Sigma}_+(\pmb{k},\omega)$ and (b) the absolute value of the imaginary part of the self-energy $|\mathrm{Im}\Sigma_+(\pmb{k},\omega)|$ as functions of $k/k_{\rm F}$ at $\omega=\xi_{\bm k}^+ $ for $\epsilon=2.5$ (red lines) and ${\epsilon=5}$ (blue lines). The inset shows the enlarged plot of $|\mathrm{Im}\Sigma_+(\pmb{k},\omega)|$ near $k=k_{\rm F}$ which is almost the same in all cases. The solid lines correspond to the modified Coulomb interaction while the dashed lines show the results obtained using the bare Coulomb interaction. The figures are plotted for $n=\rm{10^{13}~cm^{-2}}$. 
    }
\end{figure}
\section{Quasiparticle properties}
In this section we address the main quasiparticle properties defined in normal Fermi liquid formalism and report our numerical results. We consider an electron-doped monolayer of $\rm{MoS_2}$ and all the results and figures are related to the conduction band where $s=+1$. All the energies and self-energies are scaled with $\varepsilon_{\rm F}=\hbar v_{\rm F}k_{\rm F}$ and $\hbar=1$ in all calculations. The coupling constant appearing in calculations is defined as $\alpha_{\rm{ee}}=ge^2/\epsilon \hbar v_{\rm F}$ (as in graphene) with $g=4$ the band degeneracy factor and $v_{\rm F}\approx5.33\times10^{5}~\rm{m/s}$ for $\rm{MoS_2}$ and $\epsilon$ the average dielectric constant of the surrounding medium. The numerical value of the $\alpha_{\rm{ee}}$ in $\rm{MoS_2}$ is found to be $\alpha_{\rm{ee}}\approx16.4/\epsilon$.    

\subsection{Quasiparticle self-energy and spectral function}
In Fig. \ref{fig1} we plot the real and imaginary parts of the self-energy in the conduction band as functions of the scaled momentum $k/k_{\rm F}$ at the single-particle energy ${\omega=\xi_{\bm k}^+} $ and for $\epsilon=2.5$ (SiO$_2$ substrate) and $\epsilon=5$ (Al$_2$O$_3$ substrate). Here we use the on-shell approximation \cite{giuliani2005quantum} where the quasiparticle excitation energy with respect to the interacting chemical potential is given by
\begin{equation}\label{osa}
\mathcal{E}_Q^s(\bm k)\simeq \xi_{\bm k}^s+\rm{Re}\,\bar{\Sigma}_s(\pmb{k},\omega)|_{\omega=\xi_{\pmb k}^s},
\end{equation}
where $\mathrm{Re}\,\bar{\Sigma}_s(\pmb{k},\omega)=\mathrm{Re}\,\Sigma_s(\pmb{k},\omega)-\mathrm{Re} \,\Sigma_s(k_{\mathrm F},0)$ (note that the exact quasiparticle energy ${\mathcal{E}_Q^s(\bm k)= \xi_{\bm k}^s+\rm{Re}\,\bar{\Sigma}_s(\pmb{k},\omega)|_{\omega=\mathcal{E}_Q^s(\pmb k)}}$ is found from the self-consistent solving of the Dyson equation). We can see that the real part of the self-energy has a strong dip at a special momentum consistent with the strong peak in the imaginary part of the self-energy at the same momentum. This special point is the smallest momentum for which a new quasiparticle decaying channel opens in the system which is the inelastic scattering of quasiparticles into plasmon.  Bearing in mind that the quasiparticle lifetime is connected to the imaginary part of the self-energy through $\frac{1}{\tau}=-\frac{2}{\hbar}\mathrm{Im\,\Sigma}(\pmb{k},\xi_{\bm k})$, it is evident from the inset of  Fig. \ref{fig1} (b) that at $k=k_{\rm F}$, we have the most long-standing quasiparticles as expected. On the other hand, not only the wave vector at which the plasmon dip occurs changes with the interaction strength but also the nonlocal dielectric screening can abruptly push this point into smaller wave vectors. This change is more pronounced for stronger interactions or smaller dielectric constants of the surrounding medium. For $k<k_{\mathrm F}$, the behavior of the self-energy is almost the same for the bare and modified Coulomb interactions and different dielectric medium, however, the effect of the functional form of the interaction and the interaction strength come into sight as we move away from the Fermi surface and proceed toward larger wave vectors.
\begin{figure*}[htb]
\begin{center}
\subfloat      {
\includegraphics[width=.48\linewidth]{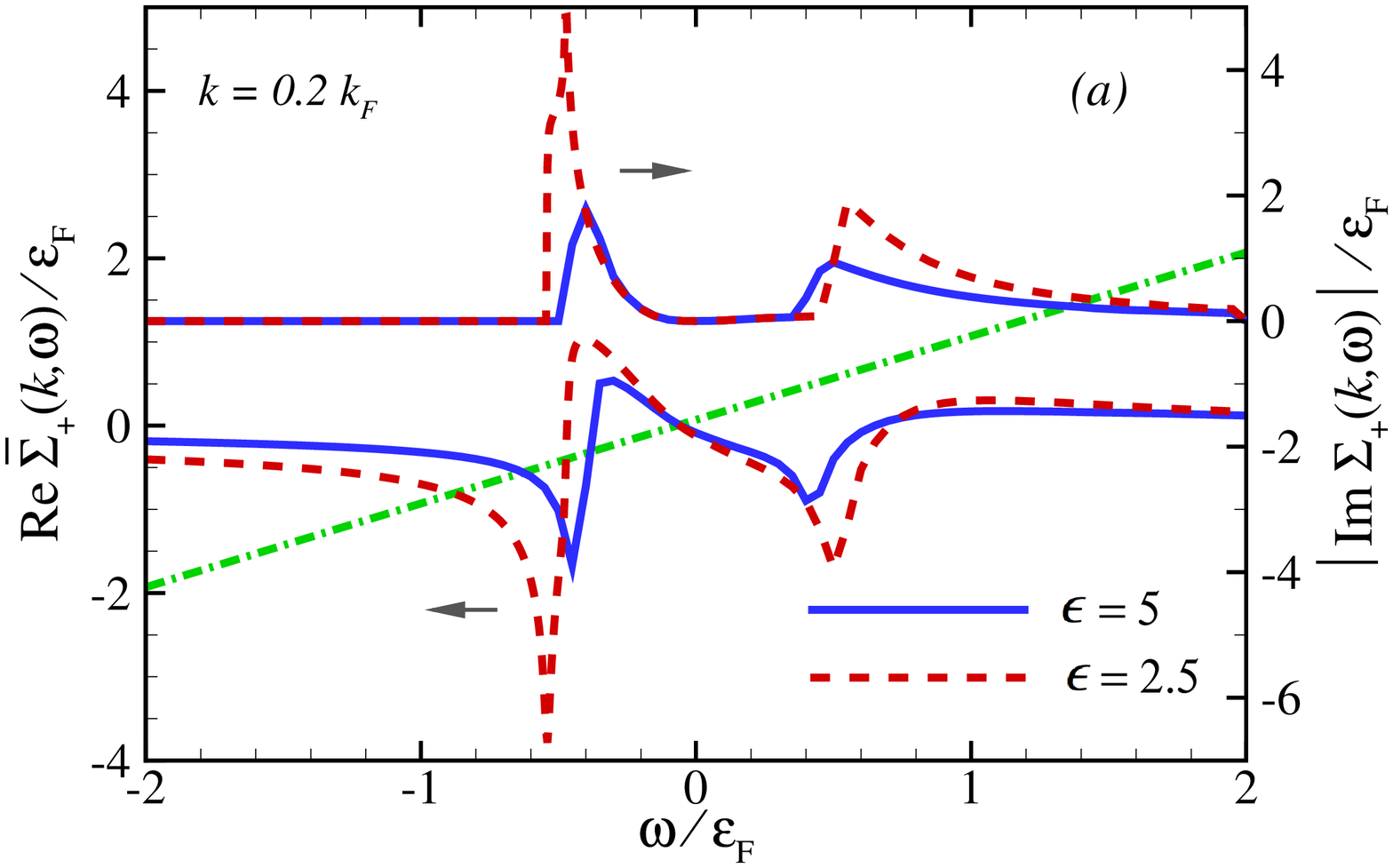}
}\quad\negthickspace\negthickspace\negthickspace\negthickspace\negthickspace\negthickspace
\subfloat   {
\includegraphics[width=.48\linewidth]{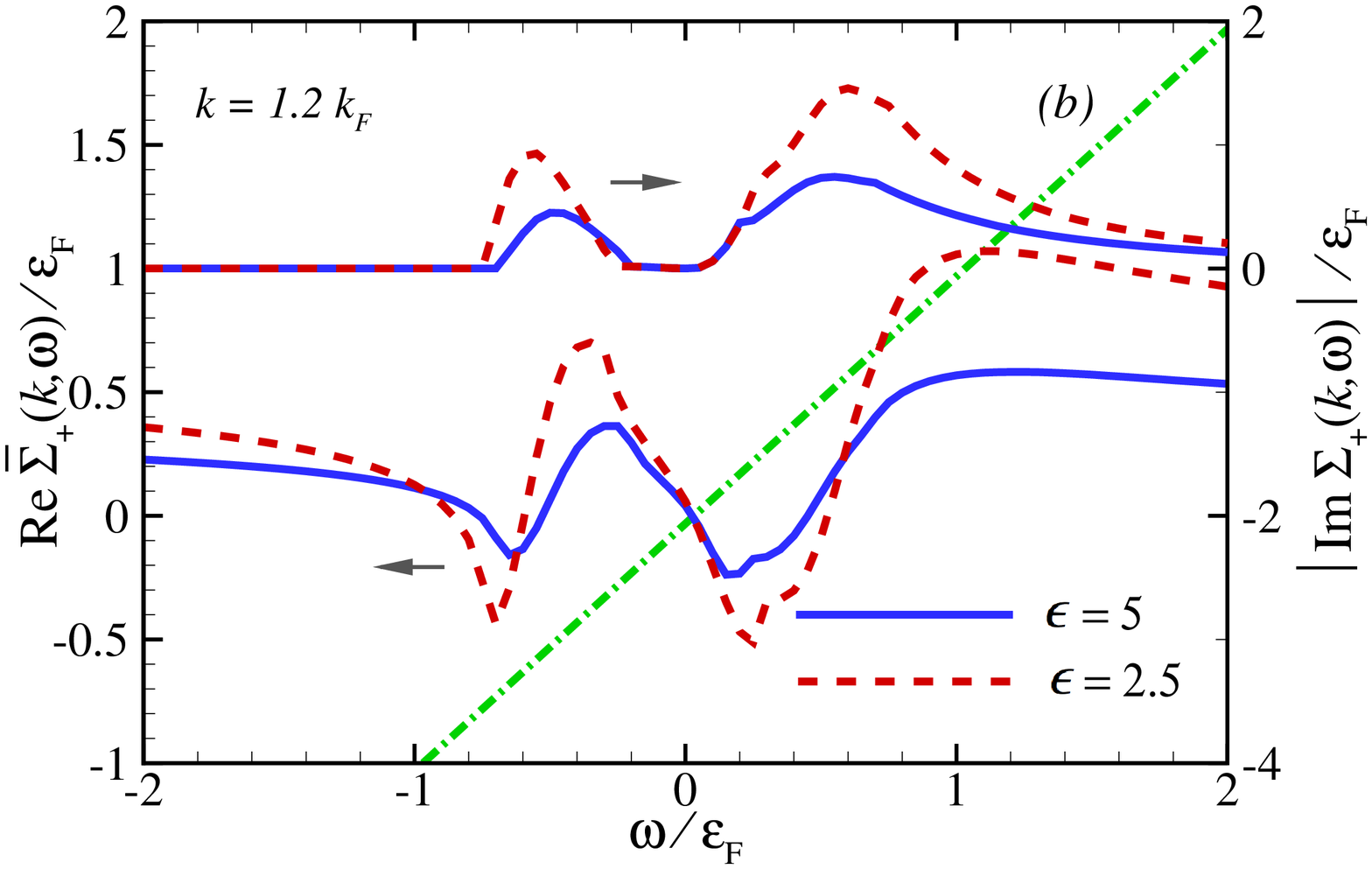}
}\\\negthickspace\negthickspace\negthickspace\negthickspace\negthickspace
\subfloat      {
\includegraphics[width=.48\linewidth]{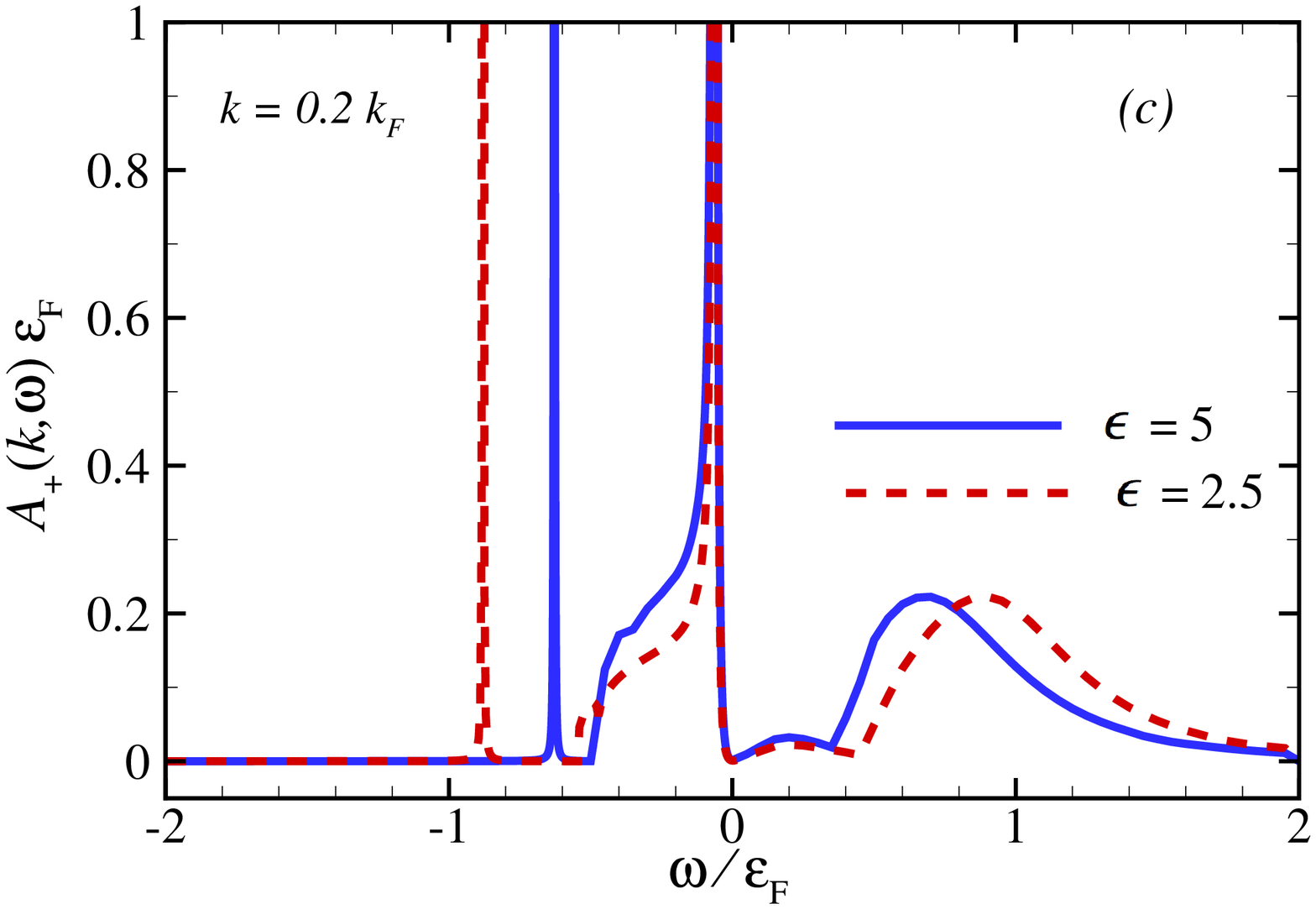}
}\quad\negthickspace\negthickspace\negthickspace\negthickspace\negthickspace\negthickspace
\subfloat    {
\includegraphics[width=.48\linewidth]{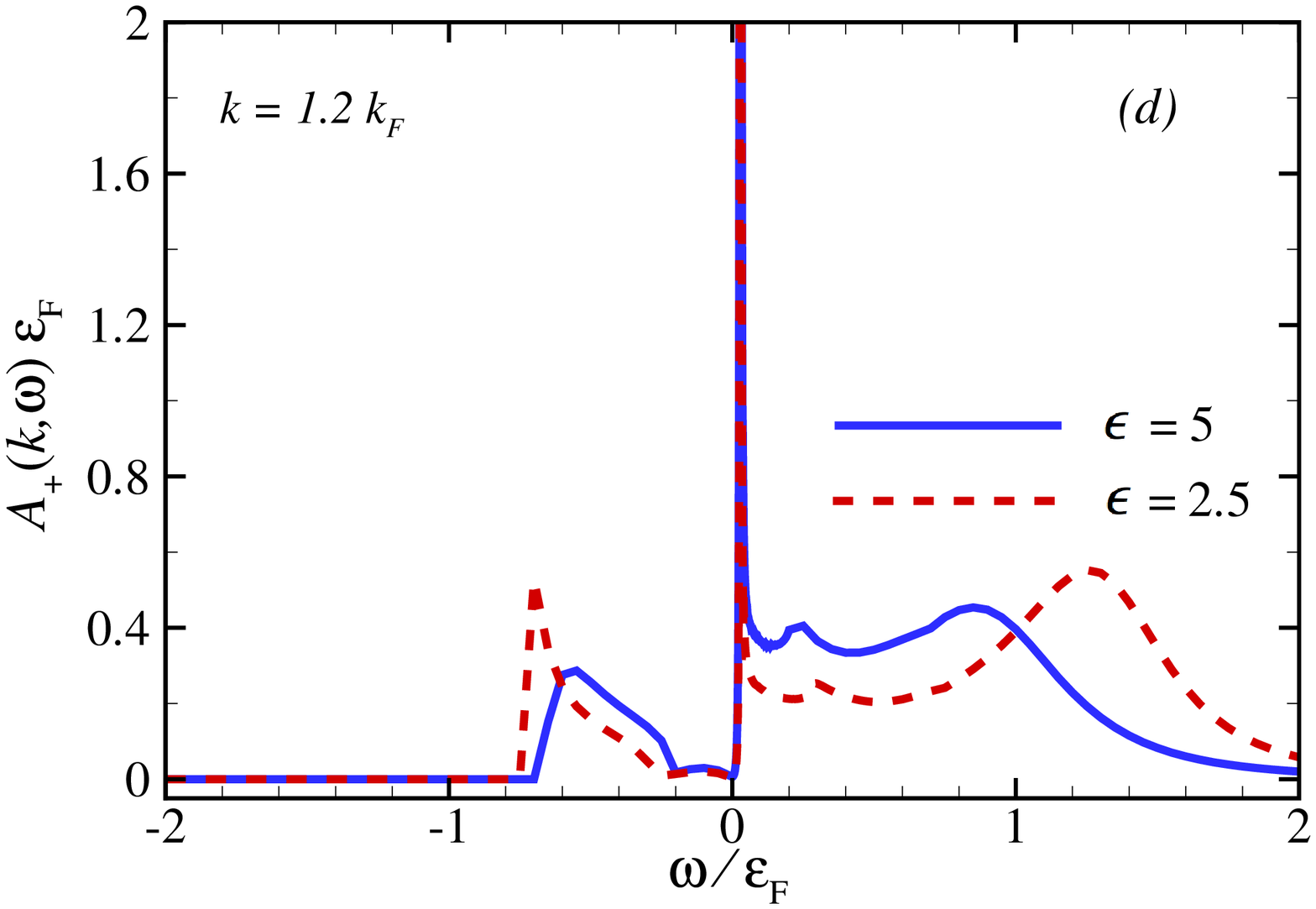}
}
\end{center}
\caption{\label{fig2} (Color online) (a),(b) The self-energy and (c),(d) the spectral function $A_+(\bm k,\omega)$ as functions of the scaled energy $\omega/\varepsilon_{\rm F}$ for $k/k_{\rm F}=0.2$ and $k/k_{\rm F}=1.2$. The dash-dotted line in (a) and (b) corresponds to $\omega -\xi^{+}_{\bm k}$ whose intersections with $\mathrm{Re}\,\bar{\Sigma}_+(\bm k,\omega)$ are solutions of the Dyson equation and show quasiparticle peaks when $\mathrm{Im}\Sigma_+(\pmb{k},\omega)$ is infinitesimal. The figures are plotted for $\epsilon=2.5$ (red dashed lines) and ${\epsilon=5}$ (blue solid lines) and $n=\rm{10^{13}~cm^{-2}}$.  }
\end{figure*}
The calculation of the real and imaginary parts of the self-energy enables us to find several quasiparticle features of the system among which the single-particle spectral function $A(\bm k,\omega)$ is of particular importance. The spectral function is a probability density function representing the probability of finding a quasiparticle with wave vector $\bm k$ and energy $\omega$ and as a probability density function, it should satisfy the sum rule ${\int_{-\infty}^{\infty}(d\omega/2\pi)A(\bm k,\omega)=1}$. For a noninteracting system, we have $A(\bm k,\omega)=2\pi\delta(w-\varepsilon_k)$ which guarantees that an excitation of the system (generated by adding or removing an electron to or from the Fermi sea) can still be described by a noninteracting particle. When the interaction is turned on, the modification of the Green's function of the system renormalizes the spectral function as \cite{mahan2013many,giuliani2005quantum}.

\begin{equation}
A_s(\bm k,\omega)=\frac{1}{\pi}\frac{|\mathrm{Im}\Sigma_s(\bm{k},\omega)|}{[\omega-\xi^{s}_{\bm k}-\mathrm{Re}\,\bar{\Sigma}_s(\bm k,\omega)]^2+[\mathrm{Im}\Sigma_s(\bm{k},\omega)]^2}
\end{equation}

The quasiparticle peaks in this case occur at ${\omega=\xi^{s}_{\bm k}+\mathrm{Re}\,\bar{\Sigma}_s(\bm k,\omega)}$ [and infinitesimal $\mathrm{Im}\Sigma_s(\bm{k},\omega)$] which are the solutions of the Dyson equation. In Fig. \ref{fig2} the typical behavior of the self-energy of the system as well as the spectral function are illustrated. The figures are plotted for two fixed values of the wave vector $k/k_{\rm F}=0.2$ and $k/k_{\rm F}=1.2$ and the modified Coulomb interaction Eq.~\eqref{vq}. The straight line is $\omega -\xi^{+}_{\bm k}$ and the intersections between this line and $\mathrm{Re}\,\bar{\Sigma}_+(\bm k,\omega)$ are solutions of the Dyson equation and represent the quasiparticles of the system provided that $\mathrm{Im}\Sigma_+(\bm{k},\omega)$ at these points is extremely small. For $k/k_{\rm F}=0.2$, we have three solutions, two of which are undamped (the first and the third one) while for the second one $\mathrm{Im}\Sigma_+(\bm k,\omega)$ is very large and therefore it has no contribution in the spectral function of the system. The first solution is the regular quasiparticle and the third one describes the plasmaron in the system which arises due to the coupling between a hole and a cloud of plasmons. For $k/k_{\rm F}=1.2$ the plasmaron peak disappears because it enters to the region where the decay process into plasmons is of much importance and the higher energy solutions are also damped leaving only the usual quasiparticle peak in the spectral function. 
On the other hand, we can see that the dielectric characteristics of the surrounding medium have the slightest impact on the usual quasiparticle peak for both values of momentum, however, for $k/k_{\rm F}=0.2$ it has considerably changed the energy of the plasmaron peak from $\omega=-0.880\,\varepsilon_{\rm F}$ for $\epsilon=2.5$ to $\omega=-0.628\,\varepsilon_{\rm F}$ for $\epsilon=5$. If we had performed our calculations with the bare Coulomb interaction instead of the modified one, the plasmaron peaks would emerge in higher energies such that for $\epsilon=2.5$, $\omega=-0.980\,\varepsilon_{\rm F}$ and  for $\epsilon=5$, $\omega=-0.681\,\varepsilon_{\rm F}$.
Along with the interaction strength and the screening length the density can also affect the quasiparticle properties. We have performed the spectral function calculations for different carrier densities and we find that changing the carrier density, the position of the regular quasiparticle peak also changes as well as that of the plasmaron peak such that for higher densities both of the peaks move to higher energies.
\subsection{The renormalization constant and the effective Fermi velocity}
In general, when the interaction is turned on, the quasiparticle peak acquires a finite width and the spectral weight reduces because of the electron-electron interaction. This reduction is of most importance at the Fermi surface and is parametrized by a renormalization constant $Z$ which is given by \cite{mahan2013many,giuliani2005quantum} 
\begin{equation}
Z=\biggl(1-\partial_{\omega}\mathrm{Re}\,{\Sigma}_+(\bm k,\omega)\bigr|_{k=k_F,\omega=0}\biggr)^{-1}.
\end{equation}

\begin{figure}[h]
\centering
\subfloat{%
  \includegraphics[width=1.\linewidth]{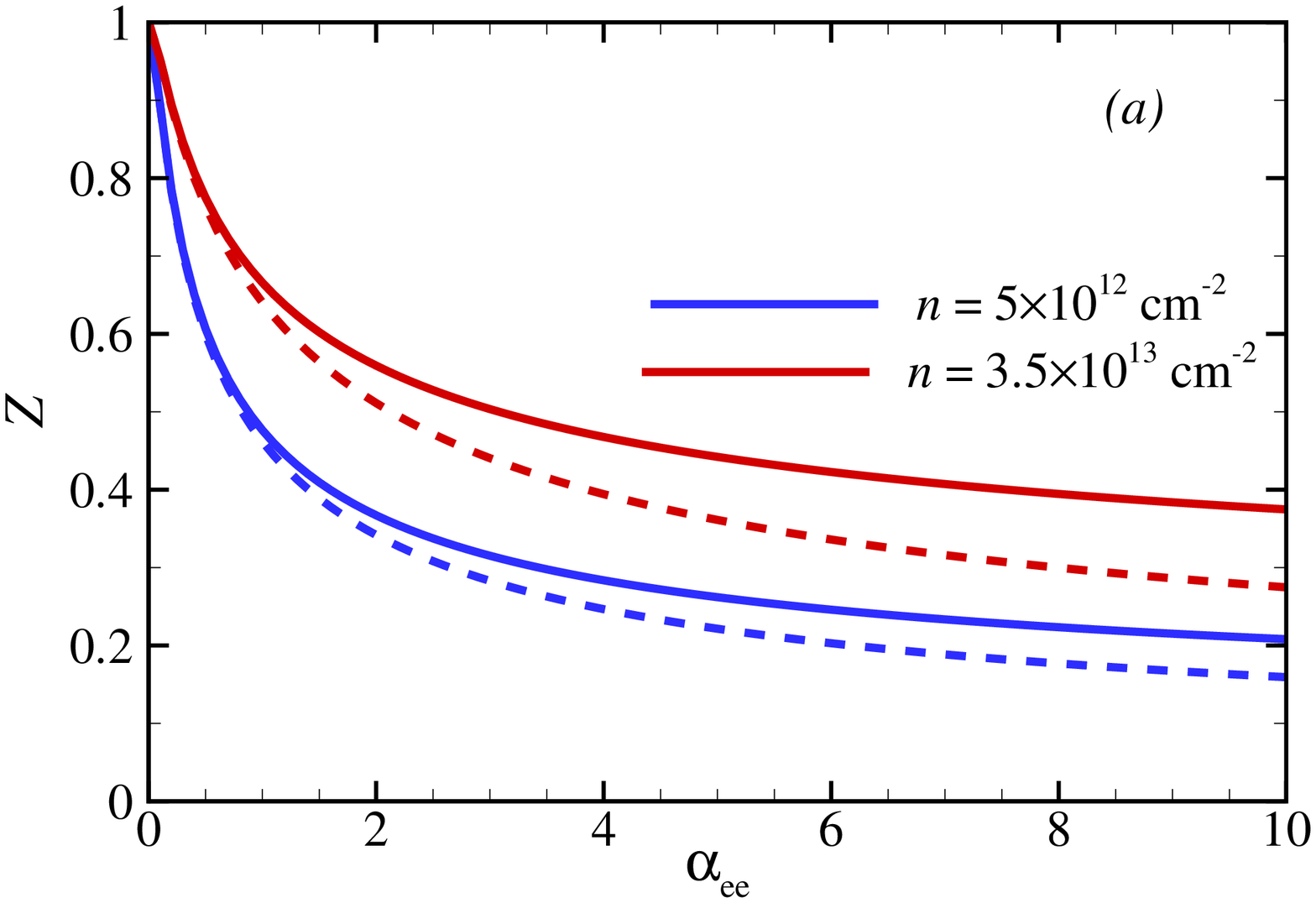}%
  }\par 
\negthickspace\negthickspace\negthickspace\negthickspace
\subfloat{%
  \includegraphics[width=1.\linewidth]{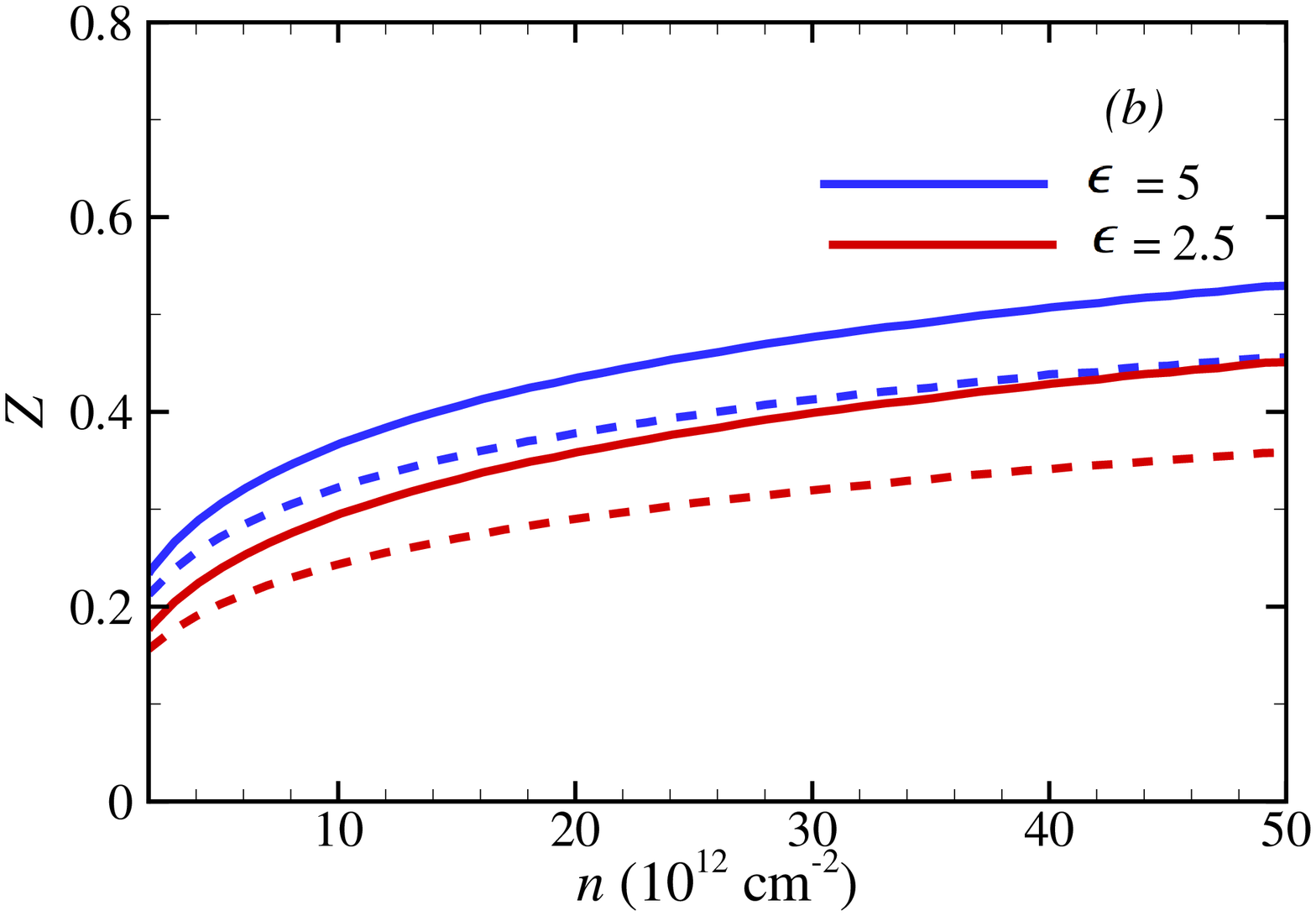}%
  }
\caption{\label{fig3} (Color online) (a) The renormalization constant $Z$ as a function of the coupling constant $\alpha_{ee}$ for ${n=\rm{3.5\times10^{13}~cm^{-2}}}$ (red lines) and $n=\rm{5\times10^{12}~cm^{-2}}$ (blue lines). (b) The renormalization constant $Z$ as a function of the electron density $n$ for $\epsilon=2.5$ (red lines) and $\epsilon=5$ (blue lines). $n$ ranges from ${\rm{2\times10^{12}~cm^{-2}}}$ to ${\rm{5\times10^{13}~cm^{-2}}}$. The solid lines correspond to the modified Coulomb interaction while the dashed lines show the results obtained using the bare Coulomb interaction.
    }
\end{figure}

It also measures the discontinuity of the occupation number at $k=k_F$ and equals to unity for a noninteracting system and $0<Z<1$ for an interacting system for which the Landau Fermi liquid picture holds. In order to satisfy the sum rule mentioned earlier, the rest of the spectral weight (given by $1-Z$) is spread incoherently in the background. In Fig. \ref{fig3} we show how the renormalization constant $Z$ varies with the electron density and also with the coupling constant $\alpha_{ee}$. We assume that the variation in coupling constant is caused by changing the surrounding environment. The variation of $Z$ versus $\alpha_{ee}$ is plotted for high and low densities $n=3.5\times 10^{13} \rm{cm^{-2}}$ and $ n=5\times10^{12} \rm{cm^{-2}}$ and also for the 
bare and modified Coulomb interactions. As expected, for $\alpha_{ee}\to 0$ the renormalization constant equals unity and the system becomes noninteracting which is the case for both high and low carrier densities. However, as we increase the coupling constant, its impact on low-density system is much more pronounced such that in the low-density case for $\alpha_{ee}=4$, we 
have $Z=0.246$ (the bare Coulomb interaction) and $Z=0.283$ (the modified Coulomb interaction) while for $n=3.5\times 10^{13} \rm{cm^{-2}}$ and at the same coupling constant we have $Z=0.468$ and $Z=0.394$ respectively. On the other hand, we can see that the inclusion of nonlocal screening increases the $Z$ factor and therefore protects the normal Fermi liquid in both 
\begin{figure}[htb]
\centering
\subfloat{%
  \includegraphics[width=1.\linewidth]{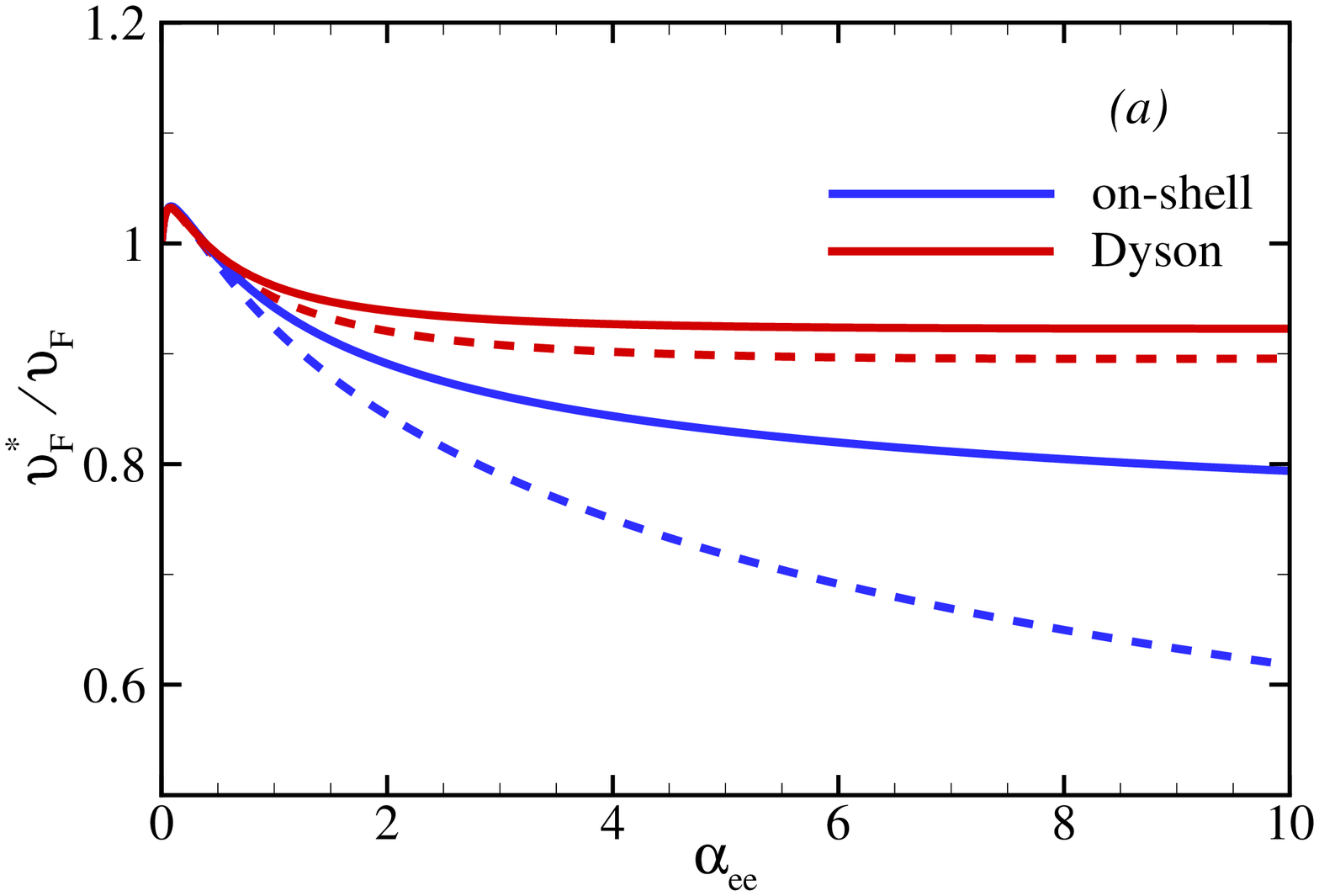}%
  }\par 
\negthickspace\negthickspace\negthickspace\negthickspace
\subfloat{%
  \includegraphics[width=1.\linewidth]{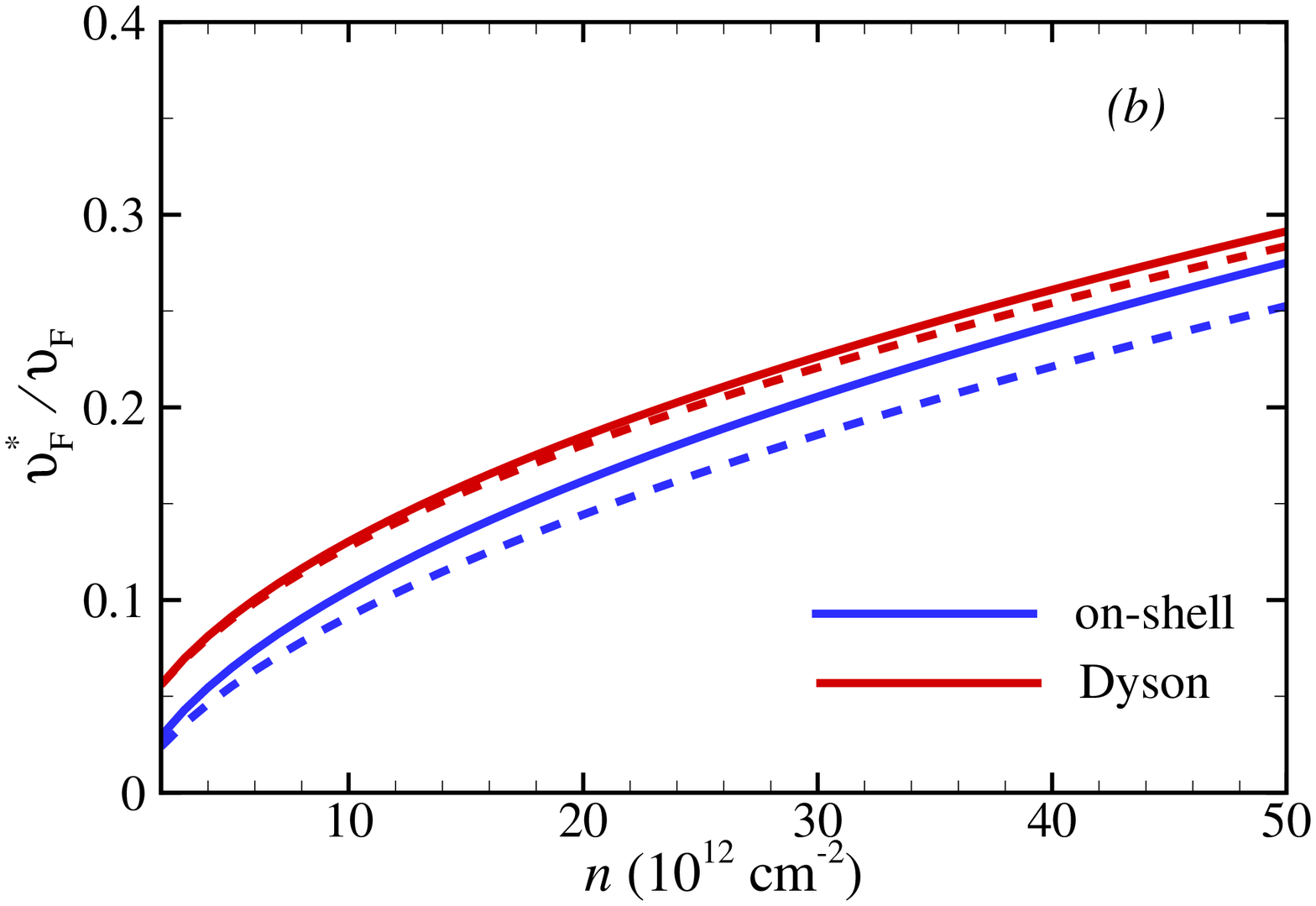}%
  }
\caption{\label{fig4} (Color online) The dimensionless effective Fermi velocity of the conduction band $v^*_{\rm F}/v_{\rm F}$ as a function of (a) the coupling constant $\alpha_{ee}$ and (b) the charge carrier density $n$ obtained using the on-shell approximation (blue lines) and the Dyson equation (red lines). ${n=\rm{3.5\times10^{13}~cm^{-2}}}$ in the top panel while $\epsilon=5$ and $n$ ranges from ${\rm{2\times10^{12}~cm^{-2}}}$ to ${\rm{5\times10^{13}~cm^{-2}}}$ in the bottom panel. The solid lines correspond to the modified Coulomb interaction while the dashed lines show the results obtained using the bare Coulomb interaction.
    }
\end{figure}
cases though the region of the effectiveness depends upon the density and the high-density system is clearly more affected. For small coupling constant (depending upon the density) the two curves of the bare and modified Coulomb interactions coincide because as we decrease the coupling constant the screening length also decreases and therefore in this limit, the bare and the modified Coulomb interactions act the same. We can also see in  Fig. \ref{fig3} (b) that increasing the density, an asymptotic value of renormalization constant is reached whose value is a function of the dielectric constant of the surrounding medium.   

In the case of massive Dirac systems the effective Fermi velocity of a quasiparticle can be defined as ${v^*_{\rm F}=\alpha |\partial \mathcal{E}_Q(\bm k)/\partial k |_{k=k_{\rm F}}}$ with $\alpha=E_{\rm F}/\varepsilon_{\rm F}$. This can be achieved by expanding the quasiparticle energy $\mathcal{E}_Q(\bm k)$ to first order in $(k-k_{\rm F})$ \cite{giuliani2005quantum} . The constant $\alpha$ is set to guarantee the equality of $v^*_{\rm F}$ with $v_{\rm F}$ when the interaction is turned off. We remember that the exact quasiparticle energy measured from the chemical potential $\mu$ of the interacting system is ${\mathcal{E}_Q^s(\bm k)= \xi_{\bm k}^s+\rm{Re}\,\bar{\Sigma}_s(\pmb{k},\omega)|_{\omega=\mathcal{E}_Q^s(\pmb k)}}$. Differentiating this equation, the effective velocity $v^{*\rm D}_{\rm F}$ (in conduction band or + channel) in the context of the Dyson equation is given by
\begin{equation}
\frac{v^{*\rm D}_{\rm F}}{v_{\rm F}}=Z\biggl(1+\frac{\alpha}{v_{\rm F}}\partial_{k}\mathrm{Re}\,{\Sigma}_+(\bm k,\omega)\bigr|_{k=k_F,\omega=0}\biggr).
\end{equation}
On the other hand the effective Fermi velocity can also be written based on the on-shell approximation  Eq.~\eqref{osa} as
\begin{equation}
\begin{split}
\frac{v^{*\rm{OS}}_{\rm F}}{v_{\rm F}}=&1+\frac{\alpha}{v_{\rm F}}\partial_{k}\mathrm{Re}\,{\Sigma}_+(\bm k,\omega)\bigr|_{k=k_F,\omega=0}\\
&+\partial_{\omega}\mathrm{Re}\,{\Sigma}_+(\bm k,\omega)\bigr|_{k=k_F,\omega=0}.
\end{split}
\end{equation}

The two definitions give the same results for very weak interactions, however, as we increase the interaction the distinction grows. There has been a long-term dispute on the validity of these approaches when an approximate form of the self-energy is employed \cite{rice1965effects,ting1975effective,lee1975landau,vinter1975correlation,asgari2005quasiparticle,zhang2005quasiparticle}. It was shown that the cancellation of higher-order corrections favors the on-shell approximation for weak interactions \cite{rice1965effects}. Even in the case of stronger interactions where the mentioned argument does not hold anymore, the two approaches are still controversial \cite{asgari2005quasiparticle,zhang2005quasiparticle}. In Fig. \ref{fig4} we compare the effective velocity found using the Dyson equation and on-shell approximation. The velocities are plotted versus the coupling constant $\alpha_{ee}$ and the electron density $n$. In Fig. \ref{fig4} (a) we can see that for weak enough interactions the two effective Fermi velocities coincide as expected. But upon increasing the interaction the effective Fermi velocity decreases in the case of the on-shell approximation (to less than $0.8\, v_{\rm F}$ for $\alpha_{ee}=10$) while at the same time $v^{*\rm D}_{\rm F}$ reaches a constant value of about $0.92\,v_{\rm F}$. Upon increasing the coupling constant to larger values, a slight upturn in $v^{*\rm D}_{\rm F}$ starts to show up (quite similar to the decline reported in $m^*$ of 2DEG at very large $r_s$ \cite{zhang2005quasiparticle}). The impact of the finite thickness of the slab is also illustrated in this figure which leads to larger effective velocities especially for the one found using on-shell approximation. This effect is suppressed for $\alpha_{ee}\to 0$ as  in the case of the $Z$ factor but increasing the coupling constant the difference between the velocities found using bare and modified interactions grows such that we have an approximately 24$\%$ increase in $v^{*\rm{OS}}_{\rm F}$ for  $\alpha_{ee}=10$ when nonlocal screening effect is considered. Note that in this figure as well as  Fig. \ref{fig3} (a), $\alpha_{ee}=6.56$ and $\alpha_{ee}=3.28$ are related to the cases where SiO$_2$ and Al$_2$O$_3$ are used as substrate. The difference between $v^{*\rm{D}}_{\rm F}$ and $v^{*\rm{OS}}_{\rm F}$ can also be noticed in Fig. \ref{fig4} (b) where we can see that the Fermi velocity is strongly suppressed at low density reminiscent of the effective mass enhancement in 2DEG. Here again, we can see the softening of the Coulomb interaction which leads to a larger effective Fermi velocity when nonlocal screening effect is considered which is stronger in the case of on-shell approximation.
 \begin{table*}[htb]
\centering
\renewcommand\arraystretch{1.3}
\caption{The calculated effective Fermi velocity and renormalization constant of $\rm{MoS_2}$ monolayer.}\label{tab1}
\begin{tabular}{@{\extracolsep{15pt}}cccccccc}
\\[-1.8ex]\hline\hline\\[-1.8ex]
{} & {} & {} & modified interaction & {} & {} & bare interaction & {} \\
\hline\\[-1.8ex]
$\alpha_{ee}$ & $n\,(10^{12}~\rm{cm^{-2}})$ & $v^{*\rm D}_{\rm F}/v_{\rm F}$ & $v^{*\rm{OS}}_{\rm F}/v_{\rm F}$ &$Z$ & $v^{*\rm D}_{\rm F}/v_{\rm F}$ & $v^{*\rm{OS}}_{\rm F}/v_{\rm F}$ &$Z$\\
\hline\\[-1.8ex]
0.2 &  $ 5 $ & 1.008 & 1.010 & 0.784 & 1.007& 1.009 & 0.782\\
{} &  $35$ & 1.021 & 1.024 & 0.894 & 1.020& 1.023 & 0.892\\
\\[-1.8ex]
1 &  $ 5 $ & 0.909 & 0.808 & 0.476 & 0.904& 0.792 & 0.462\\
{} &  $35$ & 0.961 & 0.942 & 0.666 & 0.951& 0.923 & 0.640\\
\\[-1.8ex]
3 &  $ 5 $ & 0.886 & 0.639 & 0.315 & 0.874& 0.555 & 0.283\\
{} &  $35$ & 0.931 & 0.862 & 0.503 &0.908& 0.791 & 0.440\\
\\[-1.8ex]
6 &  $ 5 $ & 0.890 & 0.554 & 0.246 & 0.874& 0.378 & 0.203\\
{} &  $35$ & 0.924 & 0.820 & 0.423 & 0.897& 0.691 & 0.336\\
\\[-1.8ex]
9 &  $ 5 $ & 0.896 & 0.517& 0.215 &0.879& 0.278 & 0.167\\
{} &  $35$ & 0.923 & 0.799 & 0.384 &0.895& 0.633 & 0.286\\
\hline

\end{tabular}
\end{table*}

We have summarized our results for the renormalization constant and effective Fermi velocity in Table~\ref{tab1} for some coupling constants and high and low densities.
\section{Summary and Conclusions}
In summary, including the dynamical effects of the electron-electron interaction, the quasiparticle properties of $\rm{MoS_2}$ monolayer are found within G$_0$W and RPA. Following the calculation of the real and imaginary parts of the self-energy in the first place, we find the dynamical structure factor of the system. We have shown that the inclusion of the nonlocal dielectric screening in the form of a modified Coulomb interaction and dielectric features of the surrounding environment have important effects on the form of the structure factor such that they can clearly change the position of the plasmaron peak, although the typical quasiparticle peak remains almost unchanged. In principle, the predicted quasiparticle peaks should be detectable in photoemission and tunneling measurements. Meanwhile, we should note that although the plasmaron peak is a long-standing theoretical prediction in condensed matter physics \cite{lundqvist1967single}, its experimental observation is rarely achieved \cite{bostwick2010observation,shay1971plasmaron,tediosi2007charge}. This is because (as our results show) the appearance and position of the plasmaron peak are very sensitive to the physical parameters of the system and a small change in the self-energy (caused for example by phonons, defects, or temperature increase) can lead to its loss in the spectral function. Besides the spectral function properties, the quasiparticle lifetime [$\tau^{-1}(\pmb{k})\sim| \mathrm{Im\,\Sigma}(\pmb{k},\xi_{\bm k}|)$] or damping rate [$\Gamma(\pmb{k})=|\mathrm{Im\,\Sigma}(\pmb{k},\xi_{\bm k})|$] and also inelastic mean free path [$l(\pmb{k})\sim k\Gamma(\pmb{k})$] are parameters of particular interest which can be tested experimentally using electron spectroscopy methods.   

We have also obtained the effective Fermi velocity and the renormalization constant of the system. The calculations are performed in a broad range of coupling constant as well as the density of the electrons in the conduction band. In the case of the renormalization constant, we find that it reduces (with respect to 1 for the noninteracting system) as the interaction increases and this reduction is more pronounced for lower densities. On the other hand, we can see that the $Z$ factor grows with the electron density and reaches an asymptotic value (depending on the surrounding environment) for very large densities which is far from unity showing that although the density of carriers is an effective parameter for the electron-electron interaction in $\rm{MoS_2}$, it can not solely describe the interacting character of the system. We recall that in the case of 2DEG, for very large values of density (or equivalently $r_s\to 0$) we have $Z\to 1$ \cite{asgari2005quasiparticle,jalabert1989quasiparticle}. The reason behind this lies in the fact that there are two scales of energy in the Hamiltonian of this system, namely $\varepsilon_{\rm F}$($\hbar v_{\rm F}k_{\rm F}$) and $\Delta$. Therefore for a constant value of the gap, the $\Delta/\varepsilon_F$ ratio is an important parameter such that for lower densities the system acts more or less like a 2DEG, but increasing the density we gradually come closer to a graphene-like system and the interaction features deviate from those of a 2DEG. 

In the case of the effective Fermi velocity, we present our results within both the self-consistent Dyson equation and the on-shell approximation. The effective Fermi velocity predicted by the on-shell approximation $v_{\rm F}^{*\rm {OS}}$  is almost always less than $v_{\rm F}^{*\rm D}$ and its behavior is completely distinct compared to $v_{\rm F}^{*\rm D}$. Increasing the coupling constant, $v_{\rm F}^{*\rm D}$ approaches a constant value not much less than the noninteracting Fermi velocity $v_{\rm F}$ while $v_{\rm F}^{*\rm {OS}}$ reduces to much smaller values (depending upon the electron density) without indicating a saturating behavior up to almost large coupling constants. In all the calculations we can trace the crucial impact of nonlocal dielectric screening leading to the softening of the Coulomb interaction and stabilizing the Fermi liquid picture especially at larger coupling constants. The predicted behavior of the effective Fermi velocity of $\rm{MoS_2}$ is experimentally testable through cyclotron resonance or Shubnikov-de Haas experiments.

To some up, our calculations show that the interaction strength of a $\rm{MoS_2}$ monolayer is not simply a function of coupling constant $\alpha_{ee}$ (as in graphene), but it is a multivariable function of $\alpha_{ee}$, $n$, and also $a$, the screening length. On this account, it is hard to comment on the accuracy of the RPA calculations, which is a leading order theory with respect to the interaction strength ($\alpha_{ee}$ for graphene and $r_s$ for 2DEG), and it is accurate for the effective interaction strength much smaller than unity. Yet, the RPA calculations have been widely used in the literature \cite{hedin1970effects,rice1965effects,dubois1959electron} for $r_s \approx 6-7$ and even larger coupling constants \cite{zhang2005quasiparticle}. In order to go beyond RPA, we need to account for the vertex corrections in the self-energy and also the dielectric function. Since a precise inclusion of the vertex terms in self-energy calculations is an unfeasibly formidable task, approximate forms of vertex corrections should be considered among which the ladder diagrams are of the most importance. Typically this is done by introducing a vertex function $\Gamma(\bm k,\omega)$ into the definitions of the self-energy and polarization function whose approximate form can be found in terms of a local field factor $G(\bm k,\omega)$ such that $\Gamma(\bm k,\omega)=1/(1+G(\bm k,\omega)V_q\chi(\bm k,\omega))$\cite{mahan2013many,giuliani2005quantum}. This approximate vertex correction is the result of replacing the average electron-electron interaction with an effective screened interaction. Obviously, the local field factor modifies the interaction in order to account for the role of exchange-correlation hole around electrons. The first and most popular local field factors were the static field factors introduced by Hubbard\cite{hubbard1957description,hubbard1958description}. But it has been shown that inclusion of the vertex corrections in the form of Hubbard-type local field factors does not change the quasiparticle properties significantly\cite{dubois1959electron,dubois1959electron2,frota1992band} and in order to find corrections to RPA calculations of quasiparticle properties, some more precise local field factors are needed which is beyond the scope of this paper. Some other improvements to this study can be achieved using a more realistic Hamiltonian for $\rm{MoS_2}$ monolayer as well as incorporating electron-phonon interaction in the electronic self-energy or performing finite temperature quasiparticle calculations.

\section{Acknowledgement}
We thank G. Vignale for fruitful discussions. This work is supported by the Iran Science Elites Federation. 
%

\end{document}